\documentclass[10pt]{article}

\usepackage[latin9]{inputenc}
\usepackage{latexsym, bm, amsmath, amssymb, graphics, amsthm,
enumerate} %
\usepackage[a4paper,
            bindingoffset=0.2in,
            left=0.8in,
            right=0.8in,
            top=0.8in,
            bottom=0.8in,
            footskip=.25in]{geometry}
\usepackage{xcolor,placeins}
\usepackage[ruled,vlined]{algorithm2e}
\usepackage[round]{natbib}
\usepackage{tikz}

\usetikzlibrary{calc,trees,positioning,arrows,chains,shapes.geometric,  decorations.pathreplacing,decorations.pathmorphing,shapes,  matrix,shapes.symbols}
\usepackage{tikz-cd}
\tikzcdset{scale cd/.style={every label/.append style={scale=#1},
    cells={nodes={scale=#1}}}}

\usepackage{hyperref} 
\usepackage{multirow,multicol}

\usetikzlibrary{arrows.meta,
                positioning,
                shapes}
\newtheorem{theorem}{Theorem}

\usepackage{comment}

\usepackage{accents}

\begin{document}

\title{Classical Simulation of Non-Classical Systems: \linebreak A Large Deviation Analysis\footnote{This paper owes a big debt to Samson Abramsky for prior joint work, to Pierre Tarres for a very careful read and important improvements, and to Vered Kurtz-David and Stuart Zoble for valuable input.  Financial support from NYU Stern School of Business, NYU Shanghai, J.P. Valles, and the HHL - Leipzig Graduate School of Management is gratefully acknowledged.  ChatGPT o3 and Grok 4 were used to help find illustrative examples and to add references.}}

\author
{Adam Brandenburger \footnote{Stern School of Business, Tandon School of Engineering, NYU Shanghai, New York University, New York, NY 10012, U.S.A., adam.brandenburger@stern.nyu.edu}
\and{Pierfrancesco La Mura \footnote{HHL - Leipzig Graduate School of Management, 04109 Leipzig, Germany, plamura@hhl.de}}
    }
\date{Version 08/07/25}
\maketitle

\thispagestyle{empty}

\begin{abstract}
Any quasi-probability representation of a no-signaling system -- including quantum systems -- can be simulated via a purely classical scheme by allowing signed events and a cancellation procedure.  This raises a fundamental question: What properties of the non-classical system does such a classical simulation fail to replicate?  We answer by using large deviation theory to show that the probability of a large fluctuation under the classical simulation can be strictly greater than under the actual non-classical system.  The key finding driving our result is that negativity in probability relaxes the data processing inequality of information theory.  We propose this potential large deviation stability of quantum (and no-signaling) systems as a novel form of quantum advantage.
\end{abstract}

\section{Introduction} \label{se1}
Bell's theorem established that no classical local hidden-variable theory can reproduce all statistical predictions of quantum mechanics (Bell, 1964).  In the decades since, an extensive theory of non-locality and contextuality has shown that quantum systems exhibit correlations incompatible with any single probability distribution over would-be measurement outcomes (see Brunner et al., 2014, and Budroni et al., 2022, for surveys).  This incompatibility can also be framed in terms of required negativity in phase-space quasi-probability representations of certain quantum states (Howard et al., 2014; Raussendorf et al., 2017).  The canonical such representation is the Wigner function (Wigner, 1932), but there are many others (Kenfack and \.Zyczkowski, 2004; Ferrie, 2011; Rundle and Everitt, 2021).  Whether couched in terms of nonlocality, contextuality, or negativity of probability, a fundamentally non-classical ingredient must appear in all examples of ``quantum advantage" in computation, communication, and cryptography (Nielsen and Chuang, 2010; Veitch et al., 2012).

The strong no-go theorems for exact classical mimicry of quantum effects notwithstanding, one may ask about approximate or resource-bounded simulations that impose classical structure and then add auxiliary resources.  Such resources include shared randomness and classical communication (Toner and Bacon, 2003; Brunner et al., 2014, Section III.C), postselection (Bacciagaluppi and Hermens, 2021), nonlocal boxes (Popescu and Rohrlich, 1994; Cerf et al., 2005), and relaxed measurement independence (Hall, 2010).  Many studies have shown that such resources come at a high quantitative cost (Brassard, Cleve, and Tapp, 1999; Massar et al., 2001; Montina, 2008; Bravyi et al., 2020).

This paper offers a new perspective on simulation cost by focusing on large deviation properties (Dembo and Zeitouni, 1998; Touchette, 2009).  We start from the observation (Abramsky and Brandenburger, 2014) that any quantum system -- indeed, any no-signaling system (Popescu and Rohrlich, 1994) -- can be simulated by a classical device that ``pushes in" negativity from probabilities to events.  In effect, the device uses ``signed events" whose plus and minus occurrences cancel out in frequency counts, thereby reproducing the average statistics of a quantum process.  This paper establishes that while this simulation succeeds at matching average statistics, it exhibits an important shortcoming.  Large deviation probabilities in the simulation may strictly exceed those in the true physical system.  Equivalently put, rare outcomes are more probable in the classical simulation than in the non-classical process itself.
 
We believe this large deviation phenomenon constitutes a new kind of quantum advantage -- to be found in the extremes of outcome statistics rather than in the mean.  Under our classical imitation of a non-classical system, large statistical fluctuations occur more frequently.  By contrast, the true physical process, precisely because it is non-classical, is able to suppress those same fluctuations more effectively.

From an information-theoretic viewpoint, the essential ingredient in our analysis is that negative probabilities can break the usual data processing inequality (e.g., Goldfeld, 2020).  Classically, if $T$ is a coarse-graining (Markov map) and $p$ and $q$ are non-negative probability measures, then $D(q || p) \geq D(Tq || Tp)$, where $D$ is the relative entropy or KL divergence.  By Sanov's theorem (Sanov, 1957), the probability of a (large) deviation to $q$ when the underlying process is $p$ is approximately $e^{-N D(q || p)}$, when the process is sampled a large number of times $N$.  Similarly, the large-$N$ probability of $Tq$ is approximately $e^{-N D(Tq || Tp)}$.  Thus, large deviations in the coarse-grained process are at least as likely as in the fine-grained process -- an expression of the basic idea of information loss from coarse-graining.

But once $T$ is a ``signed coarse-graining," as in our classical simulation, negativity can strictly reverse this inequality: $D(q || p) < D(Tq || Tp)$.  As we will see, this turns out to imply that large deviations under our simulation are more probable than under the true non-classical process.  This is the key to the quantum advantage we identify.

Historically, a clue to our direction of work can be found in Feynman's seminal writing on simulating quantum systems via a quantum computer (Feynman, 1982, 1985).  In exploring this theme, Feynman asked whether a computer operating on classical principles could simulate quantum systems.  Interestingly, he pinpointed the difficulty not in the statistical accuracy of such a scheme, but in how one could simulate negative probabilities.  Our answer is actually the reverse.  It is straightforward to simulate negative probabilities, but the statistical accuracy of the simulation -- in the large deviation sense -- is inferior.

Overall, we believe our work adds new insight into quantum reality by demonstrating that it may not be possible to mimic non-classical systems when it comes to behavior a large distance from the means of their distributions.  As well as offering foundational insight, we hope our results will lead to new ideas at the operational level of quantum mechanics.  For example, our work points to a new statistical test for non-classicality.  Testing for violations of Bell inequalities usually involves looking at expected or average correlations (Clauser et al., 1969; Aspect, Dalibard, and Roger, 1982; and, for a recent state-of the-art test, Storz et al., 2023).  Instead, one could conduct statistical tests to distinguish among probabilities of large deviations (Audenaert et al., 2007; Hiai, Mosonyi, and Ogawa, 2007).  The possible suppression of large deviations in quantum systems we have identified may also add to the known benefits of quantum over classical computing or cryptography -- applications where one might expect large departures from mean behavior to be especially costly (Tomamichel et al., 2012; Portmann and Renner, 2022).

The organization of the paper is as follows.  Section \ref{se2} briefly reviews the classical data processing inequality.  Section \ref{se3} considers a measurement-outcome model based on the Bell state (Bell, 1964) and shows that it can be simulated classically via signed events, but the data processing inequality is strictly reversed under the simulation.  Section \ref{se4} reviews in more detail how the classical simulation scheme from Abramsky and Brandenburger (2014) works.  Section \ref{se5} presents a new information-theoretic result (Theorem \ref{th1}, namely, a relaxation of the data processing inequality that reflects negativity of probability and allows reversals.  Section \ref{se6} examines the case of the uniform distribution perturbed by a small amount of negativity and proves a result (Theorem \ref{th2}) on the reversal of the data processing inequality for all sufficiently small perturbations.  Section~\ref{se7} concludes.

\section{Classical Data Processing Inequality} \label{se2}
We present the classical baseline from which our non-classical analysis of large deviations will later depart.  We illustrate the usual data processing inequality (DPI) of information theory (Goldfeld, 2020) via a simple example which is nevertheless non-trivial in that the inequality in strict.

Consider a small Ising model where we add a noisy (stochastic) measurement map in order to obtain a strict data processing gap.  There are two spins $\sigma_i \in \{+1, -1\}$, for $i = 1, 2$, with Hamiltonian $H(\sigma_1, \sigma_2) = - J\,\sigma_1\,\sigma_2$.  We set $J = 1$, so the energies are simply $E(+, +) = E(-, -) = -1$ and $E(+, -) = E(-, +) = +1$.  Setting temperature $T = 1$, we obtain the fine-grained Boltzmann (Gibbs) distribution:
\begin{align}
\nu(+, +) = \nu(-, -) = \frac{e^{+1}}{Z}, ~\label{eq1} \\
\nu(+, -) = \nu(-, +) = \frac{e^{-1}}{Z}, ~\label{eq2}
\end{align}
for partition function $Z = 2e^{+1} + 2e^{-1}$.  We next introduce noisy measurement via a Markov kernel $T$ that stochastically maps the microstates $\{++, +-, -+, -- \}$ to two macrostates $\{A, B\}$.  Let:
\begin{align}
&T_{(+,+)}(A) = 1 \;\; {\text{and}} \;\; T_{(+,+)}(B) = 0, ~\label{eq3} \\ 
&T_{(-,-)}(A) = 0 \;\; {\text{and}} \;\; T_{(-,-)}(B) = 1, ~\label{eq4} \\
&T_{(+,-)}(A) = 1/2 \;\; {\text{and}} \;\; T_{(+,-)}(B) = 1/2, ~\label{eq5} \\
&T_{(-,+)}(A) = 1/2 \;\; {\text{and}} \;\; T_{(-,+)}(B) = 1/2. ~\label{eq6}
\end{align}
The coarse-grained distribution on $\{A, B\}$ is:
\begin{align}
\mu(A) = \frac{e^{+1}}{Z} \times 1 + \frac{e^{-1}}{Z} \times 1/2 + \frac{e^{-1}}{Z} \times 1/2 = 1/2, ~\label{eq7} \\
\mu(B) = \frac{e^{+1}}{Z} \times 1 + \frac{e^{-1}}{Z} \times 1/2 + \frac{e^{-1}}{Z} \times 1/2 = 1/2. ~\label{eq8}
\end{align}

Now consider a fine-grained (large) deviation from $\nu$ to $g$ and the corresponding coarse-grained deviation from $\mu = T(\nu)$ to $f = T(g)$. By the DPI, we know that $D(g || \nu) \geq D(f || \mu)$, where $D(g || \nu)$ is the relative entropy or KL divergence:
\begin{equation}
D(g || \nu) = g(+, +)\log\frac{g(+, +)}{\nu(+, +)} + g(+, -)\log\frac{g(+, -)}{\nu(+, -)} + g(-, +)\log\frac{g(-, +)}{\nu(-, +)} + g(-, -)\log\frac{g(-, -)}{\nu(-, -)}, ~\label{eq9}
\end{equation}
and $D(f || \mu)$ is defined analogously.  If we sample the fine-grained system a large number of times $N$, then by Sanov's theorem of large deviation theory (Sanov, 1957), the probability of obtaining a frequency distribution near $g$ is approximately:
\begin{equation}
{\text{Pr}}_\nu(g; N) \approx e^{-N \times D(g || \nu)}, ~\label{eq10}
\end{equation}
and, if we sample the coarse-grained system a large number of times $N$,  the probability of obtaining a frequency distribution near $f$ is approximately:
\begin{equation}
{\text{Pr}}_\mu(f; N) \approx e^{-N \times D(f || \mu)}. ~\label{eq11}
\end{equation}
It follows that the probability of the coarse-grained deviation is weakly larger than that of the fine-grained deviation.  Intuitively, coarse-graining decreases distinguishability between distributions and therefore a deviation becomes ``less sharp" and more likely.

The DPI becomes more significant when the inequality is strict, which it is in the present example.  We use the following easily proved fact.  If $D(g || \nu) = D(f || \mu)$, then for macrostate $A$, it must be that the ratio $g(\cdot, \cdot)/\nu(\cdot, \cdot)$ is constant for all microstates that give positive probability to $A$ under $T$.  Likewise for macrostate $B$.  Microstates $(+, +), (+, -), (-, +)$ all give positive probability to $A$.  Microstates $(-, -), (+, -), (-, +)$ all give positive probability to $B$.  Therefore:
\begin{align}
{g(+, +)}/{\nu(+, +)} = {g(+, -)}/{\nu(+, -)} = {g(-, +)}/{\nu(-, +)}, ~\label{eq12} \\
{g(-, -)}/{\nu(-, -)} = {g(+, -)}/{\nu(+, -)} = {g(-, +)}/{\nu(-, +)}, ~\label{eq13}
\end{align}
from which $g = \alpha\nu$ for some constant $\alpha$.  But then, since probabilities sum to $1$, we have $\alpha = 1$, and so $g = \nu$.  By the contrapositive, whenever $g \not= \nu$, we must have $D(g || \nu) > D(f || \mu)$.  The DPI is strict and so, therefore, is the comparison of the fine-grained and coarse-grained probabilities of a deviation.

This completes our review of the classical baseline.  Next, we will see that in the context of a classical simulation of a quantum system, the DPI can be strictly reversed.

\section{Reversed Inequality for Simulation of a Quantum State} \label{se3}
We consider a measurement-outcome (``empirical") model based on the Bell state (Bell, 1964).  Alice has two possible measurements $a$ and $a^\prime$, and Bob has two possible measurements $b$ and $b^\prime$.  Associated with each measurement are two possible outcomes $0$ or $1$.  Each row in Table \ref{tb1} gives the  probabilities of the four possible outcome pairs resulting from a measurement choice by Alice and a measurement choice by Bob.

The first step in building a classical simulation of this model is to write down a phase-space representation.  To do so, we consider the set $\Omega = \{\omega_0, \omega_1, \ldots, \omega_{15}\}$.  Each $\omega_i$ encodes a $4$-bit string that specifies the outcome of each measurement in the order $a, a^\prime, b, b^\prime$.  For example, given the state $\omega_7$, we write $7 = 8^0 + 4^1 + 2^1 +1^1$ and see that this state gives outcomes $0, 1, 1, 1$ in response to measurements $a, a^\prime, b, b^\prime$, respectively.  Similarly for the other states in $\Omega$.
\begin{table}
\centering
\begin{tabular}{|c|c|c|c|c|c|}
\hline
Alice & Bob & $(0, 0)$ & $(1, 0)$ & $(0, 1)$ & $(1, 1)$ \\ \hline
$a$ & $b$ & $1/2$ & $0$ & $0$ & $1/2$ \\ \hline
$a^\prime$ & $b$ & $3/8$ & $1/8$ & $1/8$ & $3/8$ \\ \hline
$a$ & $b^\prime$ & $3/8$ & $1/8$ & $1/8$ & $3/8$ \\ \hline
$a^\prime$ & $b^\prime$ & $1/8$ & $3/8$ & $3/8$ & $1/8$ \\ \hline
\end{tabular}
\caption{Bell State} ~~\label{tb1}
\vspace{-0.2in}
\end{table}

Next, we put a probability measure $\lambda$ on $\Omega$.  We say that a phase-space probability measure realizes an empirical model if the probabilities match up in accordance with the map from states to outcomes just defined.  Take as an example the outcome pair $(0, 1)$ under the measurement pair $(a, b^\prime)$.  States $\omega_1, \omega_3, \omega_5, \omega_7$ map to this scenario.  Therefore, referring back to Table~\ref{tb1}, we require $\lambda_1 + \lambda_3 + \lambda_5 + \lambda_7 = 1/8$.  Analogous equations relate other phase-space probabilities to the remaining probabilities in the empirical model.

By Bell's theorem, we know that there is no non-negative probability measure $\lambda$ that realizes this model.  However, if we allow $\lambda$ to be a signed probability measure, that is, each $\lambda_i \in \mathbb{R}$ and $\sum_i \lambda_i = 1$, then the Bell model can be realized.  A signed $\lambda$ that works is:
\begin{multline} ~\label{eq14}
\quad\quad\quad\quad\quad\quad\quad\quad\;\;\, \lambda_0 = +1/4, \lambda_1 = +1/8, \lambda_2 = \lambda_3 = 0, \lambda_4 = +1/8, \lambda_5 = \cdots = \lambda_9 = 0, \\ \lambda_{10} = -1/8, \lambda_{11} = +1/4, \lambda_{12} = \lambda_{13} = 0, \lambda_{14} = +1/4, \lambda_{15} = +1/8. \quad\quad\quad\quad\quad\quad
\end{multline}
This is an instance of a result (Abramsky and Brandenburger, 2011, Theorem 5.9) which says that -- for general scenarios including this bipartite case -- the family of empirical models that can be realized by a signed phase-space probability measure is exactly the family of no-signaling models (Popescu and Rohrlich, 1994).  In particular, all empirical models arising from (finite-dimensional) quantum systems can be realized this way.  See the end of this section for remarks on the non-uniqueness of $\lambda$.

Our focus is on the properties of a classical simulation of $\lambda$ as given in Equation \ref{eq14}.  We follow the scheme described in Abramsky and Brandenburger (2014).  The first step is to build a doubled version of phase space, which we write as $\Psi = \Omega^+ \sqcup \, \Omega^-$ (where $\sqcup$ denotes disjoint union).  We then define a non-negative probability measure $\nu$ on $\Psi$ by:
\begin{equation} ~\label{eq15}
\nu^+_j = \begin{cases}\frac{\lambda_j}{\Lambda} \,\, \text{if} \,\, \lambda_j > 0, \\
0 \,\, \text{otherwise},
\end{cases}
\end{equation}
\begin{equation} ~\label{eq16}
\nu^-_j = \begin{cases}\frac{|\lambda_j|}{\Lambda} \,\, \text{if} \,\, \lambda_j < 0, \\
0 \,\, \text{otherwise},
\end{cases}
\end{equation}
where $\Lambda = \sum_j |\lambda_j|$.  We can now draw i.i.d.~samples from $\Psi$ under the process $\nu$.  Write:
\begin{equation} ~\label{eq17}
g = (g^+_0, \ldots, g^+_{15}, g^-_0, \ldots, g^-_{15}),
\end{equation}
for the resulting frequency distribution.  It remains to map $g$ to probabilities for the empirical model of Table \ref{tb1}.  Consider again the outcome pair $(0, 1)$ under the measurement pair $(a, b^\prime)$.  Its $g$-probability is defined as:
\begin{equation} ~\label{eq18}
\frac{(g_1^+ - g_1^-) + (g_3^+ - g_3^-) + (g_5^+ - g_5^-) + (g_7^+ - g_7^-)}{\sum_{k=0}^{15} \, (g_k^+ - g_k^-)}.
\end{equation}
That is, we calculate the net probabilities of the phase states that map to the given entry in the empirical model, where positive and negative occurrences of each state cancel, and we normalize by the total net probability.  Letting $\Gamma$ denote the map that operates as in Equation~\ref{eq18}, we note that, since it involves cancellation, this map is not a standard pushforward of probabilities.  We call $\Gamma$ a ``signed pushforward."

A quick check shows that if we set $g = \nu$ and apply $\Gamma$, we get precisely the empirical probabilities in Table~\ref{tb1}.  This justifies our calling the $\nu$-process a simulation of the phase-space probability measure $\lambda$.  It is classical because $\nu$ is non-negative.  We add further justification in the next section.

We want to assess how good the simulation $g$ is in the sense of large deviation theory.  As an example, fix the measurement pair $(a, b)$ and sample $N$ times under the probabilities in the first row of Table~\ref{tb1}, which we write as $\mu = (1/2, 0, 0 , 1/2)$.  Suppose we obtain a frequency distribution $f = (2/3, 0, 0, 1/3)$.  By Sanov's theorem, the probability of obtaining $f$ (or a frequency near $f$) is approximated for large $N$ by:
\begin{equation} ~\label{eq19}
{\text{Pr}}_\mu(f; N) \approx e^{-N \times D(f || \mu)},
\end{equation}
where $D(f || \mu) \approx 0.0566$ in the current example.  (We use $\log$ to base $e$ rather than base $2$ throughout.)  Next, choose:
\begin{equation} ~\label{eq20}
g^+_0 = 0.284, g^+_1 = 0.078, g^+_4 = 0.078, g^-_{10} = 0.170, g^+_{11} = 0.156, g^+_{14} = 0.156, g^+_{15} = 0.078,
\end{equation}
with all other components of $g$ equal to $0$.  The $g$-probability of the states where the measurement pair $(a, b)$ yields $(0, 0)$ is pushed forward under $\Gamma$ to a probability of $2/3$ -- as in the frequency distribution $f$.  The same is true for the other three components of $f$.

The probability of obtaining $g$ (or a frequency near $g$) under our classical simulation $\nu$ is approximated for large $N$ by:
\begin{equation} ~\label{eq21}
{\text{Pr}}_\nu(g; N) \approx e^{-N \times D(g || \nu)},
\end{equation}
where, using $\Lambda = 5/4$, we find:
\begin{multline} ~\label{eq22}
D(g || \nu) = 0.284\log\frac{0.284}{0.25/1.25} + 0.078\log\frac{0.078}{0.125/1.25} + 0.078\log\frac{0.078}{0.125/1.25} + 0.170\log\frac{0.170}{0.125/1.25} + \\ 0.156\log\frac{0.156}{0.25/1.25} + 0.156\log\frac{0.156}{0.25/1.25} + 0.078\log\frac{0.078}{0.125/1.25} \approx 0.0541. 
\end{multline}

We have found a strict reversal $D(g \|\nu) < D(f \|\mu)$ of the classical DPI.  It follows that in the large-$N$ limit, the probability of observing a frequency distribution near $g$ under our classical simulation is strictly higher than the probability of observing a frequency distribution near $f$ under the actual quantum process.  Formally:
\begin{equation} ~\label{eq23}
{\text{Pr}}_\nu(g; N) \approx e^{-N D(g || \nu)} \gg e^{-N D(f || \mu)} \approx {\text{Pr}}_\mu(f; N) \,\, {\text{as}} \,\, N \rightarrow \infty.
\end{equation}
The ``overshooting" in the classical simulation indicates how a quantum system can be statistically more stable against large deviations -- a consequence of its non-classical nature, which the simulation fails to replicate.

This section leaves some important questions open.  Fixing an empirical model -- such as the Bell model -- there will be typically be a many-to-one map from phase-space realizations to the given model.  The signed $\lambda$ of Equation \ref{eq14} is just one measure that realizes Table~\ref{tb1}.  This raises the question: For another realization $\lambda^\prime$ and a given row in the table, is it true or false that we can find a deviation $g^\prime$ that pushes forward to the row in question and again breaks the classical DPI?  Also open is the question of whether there is some other classical simulation that outperforms ours.  (But we do think the idea of simulation via cancellation we have examined is very natural.)  Finally, of course, there is the investigation of other quantum systems beyond the Bell model.  Our theorems in Sections \ref{se5} and \ref{se6} are a step toward a more general analysis.

\section{Simulation of Signed Probabilities} \label{se4}
We now consider the general non-classical case.  Fix a finite phase space $\Omega = \{\omega_1, \omega_2, \ldots, \omega_m\}$ and a signed probability measure $\lambda = (\lambda_1, \lambda_2, \ldots, \lambda_m)$ on $\Omega$.  We assume $\lambda_i \not= 0$ for all $i$.  Fix also a set ${\cal O} = \{o_1, o_2, \ldots, o_n\}$ of observable values and a map $\chi : \Omega \rightarrow {\cal O}$.  We push forward the signed probability measure $\lambda$ to a non-negative probability measure $\mu$ on $\cal O$ by setting:
\begin{equation} ~\label{eq24}
\mu_i = \sum_{\{j : \, \omega_j \in \chi^{-1}(o_i)\}} \lambda_j,
\end{equation}
for $i = 1, 2, \ldots, n$.  The requirement that $\mu$ is non-negative ensures that all events in $\cal O$ are observable.  Of course, for a given signed probability measure $\lambda$, this imposes a restriction on the map $\chi$.  We now draw i.i.d.~samples from $\cal O$ under the process $\mu$.  Write $f = (f_1, f_2, \ldots, f_n)$ for the resulting frequency distribution.

Following Abramsky and Brandenburger (2014), we define the classical simulation $\nu$ of $\lambda$ exactly as in the Equations \ref{eq15}-\ref{eq16} in the previous section.  As before, we draw i.i.d.~samples from $\Psi$ under the process $\nu$ and let $g$ be the resulting frequency distribution.  Also, let $\Delta(X)$ denote the usual simplex of (non-negative) probability measures on a finite set $X$.  Then the general definition of our ``signed pushforward" $\Gamma : \Delta(\Psi) \rightarrow \Delta({\cal O})$ is:
\begin{equation} ~\label{eq25}
(\Gamma(g))_i = \frac{\sum\limits_{\{j \, : \, \omega_j \in \chi^{-1}(o_i)\}} (g^+_j - g^-_j)}{\sum\limits_{k=1}^m (g^+_k - g^-_k)},
\end{equation}
for $i = 1, 2, \ldots, n$.  Heuristically, if we sample a large number of times $N$, we will obtain the approximate equality:
\begin{equation} ~\label{eq26}
(\Gamma(g))_i \approx \frac{\sum\limits_{\{j \, : \, \omega_j \in \chi^{-1}(o_i)\}} (\nu^+_j - \nu^-_j)}{\sum\limits_{k=1}^m (\nu^+_k - \nu^-_k)},
\end{equation}
so that, substituting in from Equations~\ref{eq15}-~\ref{eq16}, we get:
\begin{equation} ~~\label{eq27}
(\Gamma(g))_i \approx \frac{\sum\limits_{\{j \, : \, \omega_j \in \chi^{-1}(o_i) \, \wedge \, \lambda_j > 0\}} \lambda_j/\Lambda - \sum\limits_{\{j \, : \, \omega_j \in \chi^{-1}(o_i) \, \wedge \, \lambda_j < 0\}} |\lambda_j|/\Lambda}{\sum\limits_{\{k \, : \, \lambda_k > 0\}} \lambda_k/\Lambda - \sum\limits_{\{k \, : \, \lambda_k < 0\}} |\lambda_k|/\Lambda}.
\end{equation}
Multiplying through by $\Lambda$, we conclude:
\begin{equation} ~\label{eq28}
(\Gamma(g))_i \approx \frac{\sum_{\{j : \, \omega_j \in  \chi^{-1}(o_i)\}} \lambda_j}{1} = \mu_i,
\end{equation}
establishing that our simulation reproduces the correct probabilities.  See Abramsky and Brandenburger (2014) for the rigorous version of this argument, which uses the Strong Law of Large Numbers to obtain Equation~\ref{eq28} as a limit on a set of $\nu$-probability $1$.

This derivation justifies our definition of a signed pushforward.  Starting with a signed probability measure $\lambda$ on phase space $\Omega$, we can directly sample under the image measure $\mu$ to obtain a frequency distribution $f$ on the set of observable values $\cal O$.  Or, we can first simulate $\lambda$ via a non-negative probability measure $\nu$ on a doubled phase space $\Psi$.  We then sample under $\nu$ to obtain a frequency distribution $g$.  The transformation of $g$ via cancellation of plus-signed and minus-signed states yields an empirical distribution $\Gamma(g)$ that tends to $\mu$ as the sample size increases.

\section{Relaxation of the Data Processing Inequality} \label{se5}
In this section, we shed light on the general mechanism that allows the classical DPI to be reversed in the presence of negative probability.  For notational convenience, from this point on we will consider all distributions as defined with the phase-space indices $1, 2, \dots, m$.  To achieve this, redefine the map $\chi$ to take the index set $\{1, 2, \ldots, m\}$ into itself.  This leads to the redefinition:
\begin{equation} ~\label{eq29}
\mu_i = \sum_{\{j : \, j \in \chi^{-1}(i)\}} \lambda_j.
\end{equation}
With this, we let $f$ be the frequency distribution obtained via i.i.d.~draws under $\mu$.  Next, note that for each $j = 1, 2, \ldots, m$, either $\nu^+_j > 0$ or $\nu^-_j > 0$, not both.  So, we can re-index $\nu$ as well.  We then set $g^+_j = 0$ if $\nu^+_j = 0$ and $g^-_j = 0$ if $\nu^-_j = 0$ (these are zero-probability sampling events) and re-index $g$.  To simplify notation, we now drop the $+$ or $-$ superscripts from the $\nu_j$ and $g_j$ variables.

We can now rewrite our signed pushforward $\Gamma$ in a form that allows us to express it as a combination of a classical pushforward and a non-classical term -- analogous to the Jordan decomposition of a signed measure.  To do this, define the $m \times m$ matrix $T$ by:
\begin{equation} ~\label{eq30}
T_{ij} = \begin{cases}\text{sign}(\lambda_j) \,\, \text{if} \,\, j \in \chi^{-1}(i), \\
0 \,\, \text{otherwise}.
\end{cases}
\end{equation}
We then get the relationships:
\begin{equation} ~\label{eq31}
\mu = \Lambda \times T\nu \,\, \text{and} \,\, f = F \times Tg,
\end{equation}
where:
\begin{equation} ~\label{eq32}
F= 1/\sum_{i=1}^m (Tg)_i.   
\end{equation}

Observe that every column of $T$ contains exactly one nonzero entry, which is either $+1$ or $-1$.  Call $T$ a ``signed column-stochastic" matrix.  Following the language of information theory, we can also call $T$ a ``signed channel."  The key step is to write $T$ as the difference of two non-negative matrices: $T = T^+ - T^-$, where $T^+$ is obtained from $T$ by changing every $-1$ entry to $+1$, and $T^-$ is obtained from $T$ by changing every $+1$ entry to $0$ and every $-1$ entry to $+2$.  Note that $T^+$ is an ordinary column-stochastic matrix (a classical channel).  We calculate:
\begin{multline} ~\label{eq33}
\quad\quad\quad D(f || \mu) = \sum_i f_i \log\frac{f_i}{\mu_i} = \sum_i F(Tg)_i \log\frac{F(Tg)_i}{\Lambda(T\nu)_i} = F \sum_i (Tg)_i [\log\frac{(Tg)_i}{(T\nu)_i} + \log\frac{F}{\Lambda}] =\\ F \sum_i (Tg)_i \log\frac{(Tg)_i}{(T\nu)_i} + \log\frac{F}{\Lambda} = F \sum_i (Tg)_i \log\frac{(T^+g)_i - (T^-g)_i}{(T^+\nu)_i - (T^-\nu)_i} + \log\frac{F}{\Lambda}.
\end{multline}

We now control the departure from non-negativity.  Assume $\lambda_1 < 0$ and $\lambda_j > 0$ for all $j = 2, \ldots, m$.  It follows that $T_{i1} = -1$ for some $i$ (and all other columns contain an entry of $+1$).  From this, we can write $(T\nu)_i = -\nu_1 + K_\nu$, for some $K_\nu \geq \nu_1$.  We will further limit negativity by assuming that $|\lambda_1|$ is small, so that $\nu_1$ is small, and $ \nu_1 \ll K_\nu$.  Note that $(T^+\nu)_i = \nu_1 + K_\nu$ and $(T^-\nu)_i = 2\nu_1$.  Next write $(Tg)_i = -g_1 + K_g$.  We assume that the sample size is large, so that $g_1$ is small, and $g_1 \ll K_g$.  Note that $(T^+g)_i = g_1 + K_g$ and $(T^-g)_i = 2g_1$.  Substituting into Equation~\ref{eq33}, we get:
\begin{multline} ~\label{eq34}
\quad\quad \quad\quad D(f || \mu) = F (K_g - g_1) \log\frac{K_g - g_1}{K_\nu - \nu_1} + F \sum_{i^\prime \not=i} (T^+g)_{i^\prime}\log\frac{(T^+g)_{i^\prime}}{(T^+\nu)_{i^\prime}} + \log\frac{F}{\Lambda} = \\ F (K_g - g_1) \log\frac{K_g - g_1}{K_\nu - \nu_1} - F (K_g + g_1)\log\frac{K_g + g_1}{K_\nu + \nu_1} +  F \sum_{i^\prime=1}^m (T^+g)_{i^\prime}\log\frac{(T^+g)_{i^\prime}}{(T^+\nu)_{i^\prime}} + \log\frac{F}{\Lambda} \leq \\ F (K_g - g_1) \log\frac{K_g - g_1}{K_\nu - \nu_1} - F (K_g + g_1)\log\frac{K_g + g_1}{K_\nu + \nu_1} + F D(g || \nu) + \log\frac{F}{\Lambda},
\end{multline}
where the inequality comes from the classical DPI applied to the column-stochastic matrix $T^+$.

We now expand the first two log terms on the right hand side of Inequality~\ref{eq34} to first order and cancel resulting terms to find:
\begin{equation} ~\label{eq35}
F (K_g - g_1) \log\frac{K_g - g_1}{K_\nu - \nu_1} - F (K_g + g_1)\log\frac{K_g + g_1}{K_\nu + \nu_1} \approx F (- 2g_1\log\frac{K_g}{K_\nu} - 2g_1 + 2\frac{K_g}{K_\nu}\nu_1).
\end{equation}
Using $F = 1/(1 - 2g_1) \approx 1 + 2g_1$ and $\Lambda = 1 + 2|\lambda_1|$, we also obtain to first order:
\begin{equation} ~\label{eq36}
\log\frac{F}{\Lambda} = \log\frac{1}{(1 - 2g_1)(1 + 2|\lambda_1|)}  \approx 2g_1 - 2|\lambda_1|.
\end{equation}
Substituting Equations~\ref{eq35}-~\ref{eq36} into Inequality~\ref{eq34}, we get the approximate relation:
\begin{equation} ~\label{eq37}
D(f || \mu) \leq F(- 2g_1\log\frac{K_g}{K_\nu} - 2g_1 + 2\frac{K_g}{K_\nu}\nu_1) + (1 + 2g_1)D(g || \nu) + 2g_1 - 2|\lambda_1|.
\end{equation}

Finally, use again the large-sample approximations $g_1 \approx \nu_1$ and $K_g \approx K_\nu$, and also $\nu_1 = |\lambda_1|/\Lambda$.  Substituting into Inequality~\ref{eq37} yields the following result.

\begin{theorem} ~\label{th1}
The relative entropy $D(f || \nu)$ for the frequency distribution under the actual signed process and the relative entropy $D(g || \mu)$ for the frequency distribution under the classical simulation satisfy (to first-order for large sample sizes):
\begin{equation} ~\label{eq38}
D(f || \mu) \leq (1 + \frac{2|\lambda_1|}{\Lambda}) D(g || \nu).
\end{equation}
\end{theorem}

This is our ``signed data processing inequality" (SDPI).  As a check, note that when $\lambda_1 = 0$, we recover the classical inequality $D(f || \mu) \leq D(g || \nu)$.  The SDPI relaxes the classical version by allowing the reverse inequality $D(f || \mu) > D(g || \nu)$ to hold without contradiction.  The mechanism underlying the new inequality is the operation of a signed channel -- formally, a signed column-stochastic matrix $T$ with a column summing to $-1$.  The decomposition of $T$ into the difference of two non-negative channels $T^+$ and $T^-$ then leads to our relaxed inequality.

\section{The Near-Uniform Case} \label{se6}
In this section, we analyze a concrete family of signed probability measures for which the classical DPI is reversed.  Our approach is to inject a small amount of negativity into a uniform distribution.  We hope this is a step towards identifying more general conditions for when a signed probability measure induces a reversal.

Consider the signed probability measure:
\begin{equation} ~\label{eq39}
\lambda = (-\epsilon, \frac{1}{m} + \epsilon, \frac{1}{m}, \ldots, \frac{1}{m}),
\end{equation}
for small $\epsilon > 0$.  (For notational simplicity, we have given $\lambda$ a total of $m + 1$ components, starting at $j = 0$.)  The classical simulation $\nu$ is given by:
\begin{equation} ~\label{eq40}
\nu = (\frac{\epsilon}{1 + 2\epsilon}, \frac{1/m + \epsilon}{1 + 2\epsilon}, \frac{1/m}{1 + 2\epsilon}, \dots, \frac{1/m}{1 + 2\epsilon}).
\end{equation}
We set
\begin{equation} ~\label{eq41}
\chi(j) =
\begin{cases}
1 & \text{if} \,\, j = 0, 1, \\
j & \text{if} \,\, j = 2, \ldots, m,
\end{cases}
\end{equation}
from which:
\begin{equation} ~\label{eq42}
\mu = (0, \frac{1}{m}, \frac{1}{m}, \dots, \frac{1}{m}).
\end{equation}

Now calculate (setting $f_0 = 0$ since $\mu_0 = 0$):
\begin{align}
D(g\|\nu) &= \sum_{j=0}^m g_j \log\frac{g_j}{\nu_j} ~\label{eq43} \\
&= g_0 \log\frac{g_0}{\epsilon/(1 + 2\epsilon)} + g_1 \log\frac{g_1}{(1/m + \epsilon)/(1 + 2\epsilon)} + \sum_{j=2}^m g_j \log\frac{g_j}{(1/m)/(1 + 2\epsilon)} ~\label{eq44} \\
&=\log(1 + 2\epsilon) + g_0\log\frac{g_0}{\epsilon} + g_1\log\frac{g_1}{1/m +\epsilon} + \sum_{j=2}^m g_j \log\frac{g_j}{1/m} ~~\label{eq45} \\
&=\log(1 + 2\epsilon) + g_0\log\frac{g_0}{\epsilon} + g_1\log\frac{g_1}{1/m +\epsilon} + \sum_{j=2}^m \frac{f_j}{F} \log\frac{f_j/F}{1/m} ~~\label{eq46} \\
&= \log(1 + 2\epsilon) + g_0\log\frac{g_0}{\epsilon} + g_1\log\frac{g_1}{1/m +\epsilon} + \frac{1}{F}D(f || \mu) - \frac{f_1}{F}\log\frac{f_1}{1/m} - \frac{1 - f_1}{F}\log F. ~\label{eq47}
\end{align}
Using $F = 1/(1 - 2g_0)$ and $f_1 = F(g_1 - g_0)$, we can write:
\begin{align}
D(g\|\nu) - D(f || \mu) &= ~\label{eq48} \\
\log(1 + 2\epsilon) &+ g_0\log\frac{g_0}{\epsilon} + g_1\log\frac{g_1}{1/m +\epsilon} + (\frac{1}{F} - 1)D(f || \mu) - \frac{f_1}{F}\log(f_1m) - \frac{1 - f_1}{F}\log F = ~\label{eq49} \\
\log(1 + 2\epsilon) &+ g_0\log\frac{g_0}{\epsilon} + \bigl[g_0 + (1 - 2g_0)f_1\bigr]\log\frac{g_0 + (1 - 2g_0)f_1}{1/m +\epsilon} - 2g_0D(f || \mu) \,- ~\label{eq50} \\
&\hspace{-0.675in}(1 - 2g_0)f_1\log(f_1m) + (1 - f_1)(1 - 2g_0)\log(1 - 2g_0). ~~\label{eq51}
\end{align}

To control the behavior of $g_0\log(g_0/\epsilon)$ near $\epsilon = 0$, we set $g_0 = c\epsilon$ where $c \not= 0$, to obtain:
\begin{multline} ~\label{eq52}
D(g ||\nu) - D(f || \mu) = 
\log(1 + 2\epsilon) +c\epsilon\log c + \bigl[c\epsilon + (1 - 2c\epsilon)f_1\bigr]\log\frac{c\epsilon + (1 - 2c\epsilon)f_1}{1/m +\epsilon} \,- \\ 2c\epsilon D(f || \mu) - (1 - 2c\epsilon)f_1\log(f_1m) + (1 - f_1)(1 - 2c\epsilon)\log(1 - 2c\epsilon).
\end{multline}
It can be checked that:
\begin{equation} ~\label{eq53}
\bigl[D(g || \nu) - D(f || \mu)\bigr]\Bigm|_{\epsilon=0} = 0.
\end{equation}

Now take the derivative with respect to $\epsilon$:
\begin{multline} ~\label{eq54}
\frac{\partial [D(g ||\nu) - D(f || \mu)]}{\partial \epsilon} = \frac{2}{1 + 2\epsilon} + c\log c + (c - 2cf_1)\log\frac{c\epsilon + (1 - 2c\epsilon)f_1}{1/m + \epsilon} \, + \\
\bigl[c\epsilon + (1 - 2c\epsilon)f_1\bigr]\bigl[\frac{c -2cf_1}{c\epsilon + (1 - 2c\epsilon)f_1} - \frac{1}{1/m +\epsilon}\bigr] - 2cD(f || \mu) + 2cf_1\log(f_1m) \, + \\(1 - f_1)\bigl[-2c\log(1 - 2c\epsilon) + (1 - 2c\epsilon)\frac{-2c}{1 - 2c\epsilon}\bigr].
\end{multline}
Evaluating the derivative at $\epsilon = 0$ yields:
\begin{multline} ~\label{eq55}
\frac{\partial [D(g ||\nu) - D(f || \mu)]}{\partial \epsilon}\Bigm|_{\epsilon=0} = 2 + c\log c + c(1 - 2f_1)\log(f_1m) + c(1 - 2f_1) - f_1m \, - \\ 2cD(f || \mu) + 2cf_1\log(f_1m) - (1 - f_1)2c = 2 + c\log c + c\log(f_1m) - f_1m - 2cD(f || \mu) - c.
\end{multline}
For large sample size $N$, we can write $f_1 \approx \mu_1 = 1/m$ and $g_0 \approx \nu_0$, so that $c \approx 1$.  Using the standard non-negativity and strict positivity properties of relative entropy, we have now established the following result.

\begin{theorem} ~\label{th2}
The relative entropy $D(f || \nu)$ for the frequency distribution under the actual signed process and the relative entropy $D(g || \mu)$ for the frequency distribution under the classical simulation satisfy:
\begin{equation} ~\label{eq56}
\frac{\partial [D(g ||\nu) - D(f || \mu)]}{\partial \epsilon}\Bigm|_{\epsilon=0} \approx - 2D(f || \mu) \leq 0,
\end{equation}
where the inequality is strict if $f \not= \mu$.
\end{theorem}

It follows that if the negative component $\epsilon$ in our near-uniform probability measure $\lambda$ is small but nonzero, and we look at large-$N$ behavior, we obtain a strict reversal of the classical DPI.  We have thus identified one family of scenarios where classical simulation is always outperformed by the actual signed process.

\section{Conclusion} \label{se7}
We make some observations and then summarize.  First, a second-order Taylor series expansion yields a small-deviation version of the classical DPI (Csisz\'ar, 1967; Touchette, 2009, Section 4.2).  Fix non-negative probability measures $p$ and $q$ where $||q - p||$ is $O(1/\sqrt{N})$.  Let $M$ be a Markov kernel (coarse-graining) operating on $p$ and $q$.  Then:
\begin{equation} ~\label{eq45}
(q - p)^{\mathsf{T}} \Sigma_p^{-1} (q - p) \geq (Mq - Mp)^{\mathsf{T}} \Sigma_{Mp}^{-1} (Mq - Mp),
\end{equation}
where $\Sigma_p$ is the covariance matrix for the multinomial $p$, restricted to the simplex subspace, and $\Sigma_{Mp}$ is defined similarly.  By the Central Limit Theorem, it follows that for large $N$, the probability of the coarse-grained small deviation is at least as large as the probability of the fine-grained deviation -- a special case of the usual DPI.

The question is then whether or not the small-deviation DPI can be reversed in a non-classical setting.  We have not found a reversal in the Bell model of Section~\ref{se3}.  The reason appears to be that the signed pushforward $\Gamma$ is now applied locally rather than globally, and at this scale the map is essentially classical.  This said, we have not found a way to explore the small-deviation question in detail for non-classical systems.

The results we have been able to obtain in this paper are obviously limited -- an example of quantum advantage in the Bell model (Section~\ref{se3}), the SDPI under limited negativity (Theorem~\ref{th1}), and a family of strict reversals of the inequality again under limited negativity (Theorem \ref{th2}).  Of particular importance is to try to find more general conditions on quantum (or all no-signaling) systems under which reversals occur.  As previously mentioned, our simulation-via-cancellation scheme seems natural, but it will also be important to study other classical simulations in terms of their large deviation behavior.

Next, we distinguish our use of Sanov's theorem in this paper from the quantum Sanov theorem (Bjelakovi\'c et al., 2005, N\"otzel, 2014), which is an extension of the classical Sanov theorem to the quantum setting using relative von Neumann entropy.  By contrast, we use only the classical Sanov theorem because, while we consider quantum settings, our interest is in the properties of a classical simulation.

In conclusion, we have found that the role of negativity of probability in non-classical systems goes beyond violating Bell-type inequalities or providing computational speed-ups.  It may also endow these systems with a statistical robustness against large fluctuations, as compared with a classical simulation.  We propose this robustness as a new way in which negativity acts as core resource in non-classical physics.

\section*{References}

\noindent Abramsky, S., and A. Brandenburger, ``The Sheaf-Theoretic Structure of Non-Locality and Contextuality," \textit{New Journal of Physics}, 13, 2011, 113036.

\vspace{0.1in}

\noindent Abramsky, S., and A. Brandenburger, ``An Operational Interpretation of Negative Probabilities and No-Signalling Models," in van Breugel, F., E. Kashefi, C. Palamidessi, and J. Rutten (eds.), \textit{Horizons of the Mind: A Tribute to Prakash Panagaden}, Lecture Notes in Computer Science 8464, Springer, 2014, 59-75.

\vspace{0.1in}

\noindent Aspect, A., J. Dalibard, and G. Roger, ``Experimental Test of Bell's Inequalities Using Time-Varying Analyzers," \textit{Physical Review Letters}, 49, 1982, 1804.

\vspace{0.1in}

\noindent Audenaert, K., J. Calsamiglia, R. Mu\~noz-Tapia, E. Bagan, Ll. Masanes, A. Acin, and F. Verstraete, ``Discriminating States: The Quantum Chernoff Bound," \textit{Physical Review Letters}, 98, 2007, 160501.

\vspace{0.1in}

\noindent Bacciagaluppi, G., and R. Hermens, ``Bell-Inequality Violation and Relativity of Pre- and Postselection," \textit{Physical Review A}, 104, 2021, 012201.

\vspace{0.1in}

\noindent Bell, J., ``On the Einstein-Podolsky-Rosen Paradox," 1, \textit{Physics}, 195-200.

\vspace{0.1in}

\noindent Bjelakovi\'c, I., J.-D. Deuschel, T. Kr\"uger, R. Seiler, R. Siegmund-Schultze, and A. Szkola, ``A Quantum Version of Sanov's Theorem," \textit{Communications in Mathematical Physics}, 260, 2005, 659-671.

\vspace{0.1in}

\noindent Brassard, G., R. Cleve, and A. Tapp, ``Cost of Exactly Simulating Quantum Entanglement with Classical Communication," \textit{Physical Review Letters}, 83, 1999, 1874. 

\vspace{0.1in}

\noindent Bravyi, S., D. Gosset, R. K\"onig, and M. Tomamichel, ``Quantum Advantage with Noisy Shallow Circuits," \textit{Nature Physics}, 16, 2020, 1040-1045.

\vspace{0.1in}

\noindent Brunner, N., D. Cavalcanti, S. Pironio, V. Scarini, and S. Wehner, ``Bell Nonlocality," \textit{Reviews of Modern Physics}, 86, 2014, 419-478.

\vspace{0.1in}

\noindent Budroni, G., A. Cabello, O. G\"uhne, M. Kleinmann, and J.-\r{A}. Larsson, ``Kochen-Specker Contextuality," \textit{Reviews of Modern Physics}, 94, 2022, 045007.

\vspace{0.1in}

\noindent Cerf, N., N. Gisin, S. Massar, and S. Popescu, ``Simulating Maximal Quantum Entanglement without Communication," \textit{Physical Review Letters}, 94, 2005, 220403.

\vspace{0.1in}

\noindent Clauser, J., M. Horne, A. Shimony, and R. Holt, ``Proposed Experiment to Test Local Hidden-Variable Theories," \textit{ Physical Review Letters}, 23, 1969, 880.

\vspace{0.1in}

\noindent Csisz\'ar, I., ``Information-Type Measures of Difference of Probability Distributions and Indirect Observations," \textit{Studia Scientiarum Mathematicarum Hungarica}, 2, 1967, 299-318.

\vspace{0.1in}

\noindent Dembo, A., and O. Zeitouni, \textit{Large Deviations: Techniques and Applications}, Springer, 2nd edition, 1998.

\vspace{0.1in}

\noindent Ferrie, C., ``Quasi-Probability Representations of Quantum Theory with Applications to Quantum Information Science," \textit{Reports on Progress in Physics}, 74, 2011, 116001.

\vspace{0.1in}

\noindent Feynman, R., ``Simulating Physics with Computers," \textit{International Physics with Computers}, 21, 1982, 467-488.

\vspace{0.1in}

\noindent Feynman, R., ``Quantum Mechanical Computers," \textit{Optics News}, 11, 1985, 11-20.

\vspace{0.1in}

\noindent Goldfeld, Z., ``Information Theory for Data Transmission, Security, and Machine Learning," ECE 5630, Cornell University, 2020, Lecture 7, at http://people.ece.cornell.edu/zivg/ECE\_5630\_Lectures7.pdf.

\vspace{0.1in}

\noindent Hall, M., ``Local Deterministic Model of Singlet State Correlations Based on Relaxing Measurement Independence," \textit{Physical Review Letters}, 105, 2010, 250404.

\vspace{0.1in}

\noindent Hiai, F., M. Mosonyi, and T. Ogawa, ``Large Deviations and Chernoff Bound for Certain Correlated States on a Spin Chain," \textit{Journal of Mathematical Physics}, 48, 2007, 123301.

\vspace{0.1in}

\noindent Howard, M., J. Wallman, V. Veitch, and J. Emerson, ``Contextuality Supplies the Magic for Quantum Computation," \textit{Nature}, 510, 2014, 351-355.

\vspace{0.1in}

\noindent Kenfack, A., and K. \.Zyczkowski, ``Negativity of the Wigner Function as an Indicator of Non-Classicality," \textit{Journal of Optics B: Quantum and Semiclassical Optics}, 6, 2004, 396-404.

\vspace{0.1in}

\noindent Massar, S., D. Bacon, N. Cerf, and R. Cleve, ``Classical Simulation of Quantum Entanglement Without Local Hidden Variables," \textit{Physical Review A}, 63, 2001, 052305.

\vspace{0.1in}

\noindent Montina, A., ``Exponential Complexity and Ontological Theories of Quantum Mechanics, \textit{Physical Review A}, 77, 2008, 022104.

\vspace{0.1in}

\noindent Nielsen, M., and I. Chuang, \textit{Quantum Computation and Quantum Information}, Cambridge University Press, 2010.

\vspace{0.1in}

\noindent N\"otzel, J. ``Hypothesis Testing on Invariant Subspaces of the Symmetric Group: Part I. Quantum Sanov's Theorem and Arbitrarily Varying Sources," \textit{Journal of Physics A: Mathematical and Theoretical}, 47, 2014, 235303.

\vspace{0.1in}

\noindent Popescu, S., and D. Rohrlich, ``Quantum Nonlocality as an Axiom," \textit{Foundations of Physics}, 24, 1994, 379-385.

\vspace{0.1in}

\noindent Portmann, C., and R. Renner, ``Security in Quantum Cryptography," \textit{Reviews of Modern Physics}, 94, 2022, 025008.

\vspace{0.1in}

\noindent Raussendorf, R., N. Delfosse, D. Browne, C. Okay, and J. Bermejo-Vega, ``Contextuality and Wigner-Function Negativity in Qubit Quantum Computation," \textit{Physical Review A}, 95, 2017, 052334.

\vspace{0.1in}

\noindent Rundle, R., and M. Everitt, ``Overview of the Phase Space Formulation of Quantum Mechanics with Application to Quantum Technologies," \textit{Advanced Quantum Technologies}, 4, 2021, 2100016.

\vspace{0.1in}

\noindent Sanov, I., ``On the Probability of Large Deviations of Random Variables," \textit{Mat. Sbornik}, 42, 1957, 11-44.

\vspace{0.1in}

\noindent Storz, S., J. Sch\"ar, A. Kulikov, P. Magnard, P. Kurpiers, J. L\"utolf, T. Walter, A. Copetudo, K. Reuer, A. Akin, J.-C. Besse, M. Gabureac, G. Norris, A. Rosario, F. Martin, J. Martinez, W. Amaya, M. Mitchell, C. Abellan, J.-D. Bancal, N. Sangouard, B. Royer, A. Blais and A. Wallraff, ``Loophole-Free Bell Inequality Violation with Superconducting Circuits, \textit{Nature}, 617, 2023, 265-270.

\vspace{0.1in}

\noindent Tomamichel, M., C. Lim, N. Gisin, and R. Renner, ``Tight Finite-Key Analysis for Quantum Cryptography," \textit{Nature Communications}, 3, 2012, 634.

\vspace{0.1in}

\noindent Toner, B., and D. Bacon, ``Communication Cost of Simulating Bell Correlations," \textit{Physical Review Letters}, 91, 2003, 187904.

\vspace{0.1in}

\noindent Touchette, H., ``The Large Deviation Approach to Statistical Mechanics," \textit{Physics Reports}, 478, 2009, 1-69.

\vspace{0.1in}

\noindent Veitch, V., C. Ferrie, D. Gross, and J. Emerson, ``Negative Quasi-Probability as a Resource for Quantum Computation," \textit{New Journal of Physics}, 14, 2012, 113011.

\vspace{0.1in}

\noindent Wigner, E., ``On the Quantum Correction for Thermodynamic Equilibrium," \textit{Physical Review}, 40, 1932, 749-759.

\end{document}